\documentclass[onecolumn,preprintnumbers]{revtex4}
\usepackage{amsmath,amssymb,graphics,epsfig,subfigure}
\usepackage{color}
\usepackage[colorlinks,linkcolor=red,anchorcolor=red,citecolor=green]{hyperref}
\usepackage{setspace}
\usepackage{booktabs}
\usepackage{float}

\newcommand{\red}[1]{{\color{red} #1}}
\newcommand{\green}[1]{{\color{green} #1}}

\begin{document}

\thispagestyle{empty}

\begin{center}

\title{van der Waals fluid and charged AdS black hole in the Landau theory}

\author{Zhen-Ming Xu \footnote{E-mail: xuzhenm@nwu.edu.cn}, Bin Wu, and Wen-Li Yang
        \vspace{6pt}\\}

\affiliation{ $^{1}$Institute of Modern Physics, Northwest University, Xi'an 710127, China\\
$^{2}$School of Physics, Northwest University, Xi'an 710127, China\\
$^{3}$Shaanxi Key Laboratory for Theoretical Physics Frontiers, Xi'an 710127, China\\
$^{4}$Peng Huanwu Center for Fundamental Theory, Xi'an 710127, China}

\begin{abstract}
By introducing the general construction of Landau functional of the van der Waals system and charged AdS black hole system, we have preliminarily realized the Landau continuous phase transition theory in black hole thermodynamics. The results show that the Landau functional constructed in present paper can directly reflect the physical process of black hole phase transition. Specifically, the splitting of the global minimum of the Landau functional corresponds to the second-order phase transition of the black hole, and the transformation of the global minimum reflects the first-order phase transition of the black hole.
\end{abstract}

\maketitle
\end{center}

\section{Introduction}
Since Hawking and Bekenstein proposed the entropy and temperature of the black hole~\cite{Bekenstein1973,Bardeen1973}, after the development for nearly 50 years, black hole thermodynamics has become an indispensable tool for the study of black hole physics and even quantum gravity, and shows a strong vitality. The most typical is the Hawking-Page phase transition~\cite{Hawking1983}, which now is interpreted as the confinement/deconfinement phase transition in gauge theory~\cite{Witten1998}. Then the study of black hole thermodynamics extends to some AdS black holes~\cite{Chamblin1999a,Chamblin1999b,Romans1992}. Nowadays, with the introduction of extended phase space~\cite{Kastor2009,Dolan2011a,Dolan2011b}, the concept of black hole thermo-physics has become more abundant, such as van der Waals fluid phase transition~\cite{Kubiznak2012}, black hole holographic heat engine~\cite{Johnson2014a}, black hole chemistry~\cite{Kubiznak2017}, black hole thermodynamics geometry~\cite{Ruppeiner2014,Wei2015,Miao2018,Wei2019,Xu2020a,Xu2020b,Ghosh2020,Wei2020a}, quantum phase transition~\cite{Anabal2021a,Anabal2021b} and so on~\cite{Hendi2016,Hendi2019,Bhattacharya2017,Guo2019,Caceres2015}. On the thermodynamic phase transition of the black hole, we have known that the Maxwell equal area law and Gibbs free energy of the system can accurately give the phase diagram of the black hole. Recently, some studies have explored the evolution process of black hole phase transition from the perspective of kinetics, based on the so-called free energy landscape, using the Fokker-Planck equation in non-equilibrium statistical physics~\cite{Li2020a,Li2020b,Wei2020b,Li2020c}. All these studies provide a good guidance for us to understand the microscopic information of black holes.

Phase transition is the core of a thermodynamic system. Understanding some physical processes or the dynamic evolution of phase transition has become an urgent problem. In retrospect, people mostly focused on the analytical calculation of the phase transition and the criticality, and often ignored some physical connotations in this process. The free energy landscape is a good research project~\cite{Li2020a,Li2020b,Wei2020b,Li2020c}. Based on some non-equilibrium means, we can explore the kinetics of evolution between different phases of the black hole. We know that black hole thermodynamics is a kind of equilibrium thermodynamics, so we need to ask that from the perspective of equilibrium, how to understand the evolution process of black hole phase transition, or how to intuitively obtain some physical mechanisms in the process of all phase transition at least in spirit if not in detail. In this paper, we are trying to explore this problem, which is also the main motivation of our present paper.

With the help of Landau theory, we construct the Landau functional, or the Landau free energy, of a thermodynamic system. By introducing an auxiliary parameter, we can intuitively obtain the thermodynamic phase transition process of the system (mainly van der Waals system and charged AdS black hole system). To sum up, the splitting of the local minimum of the Landau functional reflects the second-order phase transition of a thermodynamic system, while the transformation of the global minimum corresponds to the first-order phase transition. Although the newly introducing procedure is too simplistic, in principle the idea of having a Landau functional in the extended phase space is a good one, and worth of elaborating. All in all, this scheme can provide a new direction for us to study the microscopic information of a thermodynamic system, especially for the black hole system in the future.

\section{Landau functional}
Landau approximated the free energy of a system, in such a way that it exhibits the non-analyticity of a phase transition and turns out to capture much of the physics. Landau free energy $L$, or sometimes the Landau functional, has some characteristics~\cite{Goldenfeld1992}.
\begin{itemize}
  \item Landau functional $L$ has the property that the state of the system is specified by the global minimum of $L$ with respect to the order parameter.
  \item Landau functional $L$ has dimension of energy, and is related to, but is not identical with the Gibbs free energy of the system.
\end{itemize}

For a general thermodynamic system, we can construct the following Landau functional
\begin{eqnarray}\label{landau}
L=\int F(X,T,P)\text{d}X
\end{eqnarray}
where the parameter $X$ is treated as an auxiliary variable, while the temperature $T$ and pressure $P$ of the thermodynamic system are treated as independent parameters. The function $F(X,T,P)$ represents some relations satisfied by these three important parameters in the whole thermodynamic system.

The equation of state for the thermodynamic system can be written as $P=f(V,T)$, where $V$ is thermodynamic volume of the thermodynamic system. According to the equation of state, as a preliminary and coarse-grained description, we can construct the function $F(X,T,P)$ by working backward
\begin{eqnarray}\label{landauf}
F(X,T,P)\equiv P-f(X,T).
\end{eqnarray}

For black holes, the thermodynamic pressure $P$ is just related to the AdS radius, while other variables (like $T$, $S$ and $V$) are the functions of the event horizon radius and other parameters. Thus the pressure is a completely independent and special quantity. We know that for a simple thermodynamic system, the equation of state $F(V,T,P)=0$ is a thermodynamic equation describing the state of the system under a set of given physical conditions. There may be many different paths in the process of the system reaching the equilibrium state, and the most preferred path of the system is the one that makes the free energy of the system take the minimum value. In the isothermal and isobaric environment, we introduce an auxiliary variable $X$ with the dimension of volume, into the process of the system approaching equilibrium. One thing that can be determined is that through this parameter, we can obtain some real physical processes of a thermodynamic system. In this sense, this parameter acts as an order parameter.

In this way, among all the possible relations $F(X,T,P)$ satisfied by the three parameters $\{X,T,P\}$, the state that can make the functional $L$ take the minimum value is the most {\em real} state of the system, i.e.,
\begin{eqnarray}
\frac{\text{d}L}{\text{d}X}=F(X,T,P)=0 \quad \Rightarrow \quad X=V,
\end{eqnarray}
which exactly corresponds to the state of a thermodynamic system, i.e., $F(V,T,P)=0$. Now we try to understand the physical meaning of the newly introduced order parameter $X$. It is a certain volume of the system in the process of reaching equilibrium, or not strictly speaking, can be considered as ``the volume of the system in the non-equilibrium state'', which does not satisfy the equation of state of the system. While the thermodynamic volume $V$ in equilibrium is the root of the function $F(X,T,P)=0$ and it satisfies the equation of state of the system. Moreover, another advantage of the Landau functional constructed in this way is that the concavity and convexity of the extreme point are related to the thermal stability of the thermodynamic system,
\begin{eqnarray}
\delta\left(\left.\frac{\text{d}L}{\text{d}X}\right|_{X=V}\right)=\frac{\partial f(V,T)}{\partial V}\delta V,
\end{eqnarray}
from which we can clearly see that when $\partial f(V,T)/\partial V >0$, the extreme point is similar to a potential well, and the corresponding thermodynamic state is stable, while when $\partial f(V,T)/\partial V <0$, the extreme point is similar to a potential barrier, and the corresponding thermodynamic state is unstable.

Next, we will make further comments on the above construction scheme.
\begin{itemize}
   \item The parameter $X$ has the dimension of volume. If $X=V$, according to the first law of thermodynamics $dE=TdS-PdV$, where $E$ is the internal energy, we can find $L=E+PV-TS$, which is the Gibbs free energy (on-shell). For general $X$, we can obtain $L=PX-\int f(X,T)dX$. It has dimension of energy, and can be considered as (off-shell) free energy, whose concrete form depends on the expression of $f(X,T)$.
  \item Our current method is relatively elementary, and is based on the equation of state of the system. We treat it as an inverse problem, thinking that the equation of state should be the solution of a certain function. Simply considering that the equation of state corresponds to the pole of a function, we have the construction of Eq.~(\ref{landau}) in this paper. Therefore, some thermodynamic properties (thermal stability) are directly reflected in some geometric characteristics (concavity and convexity) at the extreme points of the constructed functional.
  \item In the case of usual free energy computation, the thermodynamic ensemble is fixed by the boundary conditions which make the action well defined. In that way it is clear how a change of the boundary conditions produces a Legendre transformation in the free energy. For our case, at present, we have not involved in this basic and important issue. We are only considering an inverse problem. It is a bottom-up process to deduce the possible sources according to the existing thermodynamic behaviors of the system, and some basic problems involved in this process need to be solved in the future.
  \item Compared to the usual $P-V$ criticality analysis, the biggest advantage of this method is that it can dynamically demonstrate the phase transition process, which provides a premise for us to analyze some details of phase transition. We can use the stochastic process to analyze the dynamics of phase transition, just like the free energy landscape in the literature~\cite{Li2020a,Li2020b,Wei2020b,Li2020c}. At present, the disadvantages are as follows: first, for more complex black holes (such as Kerr black hole), due to the complexity of the thermodynamic volume, it is not so easy to get the exact expression of Landau functional; second, calculation process of the critical exponent for phase transition from the current Landau functional may fall back to the usual $P-V$ criticality analysis, while the known Landau functional obtained from the Euclidean action can directly give the critical exponent through its related coefficients.
\end{itemize}

With the Landau functional, we can more intuitively understand the microscopic mechanism of phase transitions of the thermodynamic system. Next we use two specific examples to visualize the current general discussion.

\section{For van der Waals fluid}
The equation of state for the van der Waals fluid reads as~\cite{Johnston2014b}
\begin{eqnarray}\label{eosvdw}
\left(P+\frac{a}{v^2}\right)(v-b)=\tau,
\end{eqnarray}
where $P$ is pressure, $\tau=k_{B}T$ and $k_{B}$ is the Boltzmann constant, and $T$ is temperature. The constant $b > 0$ is a measure of the size of the molecules, whereas the constant $a > 0$ is related to the attraction between the molecules. The specific volume of the van der Waals fluid is $v=V/N$, where $V$ is the volume occupied by the fluid and $N$ is the number of molecules. We can write the chemical potential $\mu$, or equivalently the Gibbs free energy $G=\mu N$~\cite{Johnston2014b},
\begin{eqnarray}\label{chemical}
\mu=-\tau\ln(v-b)+\frac{b\tau}{v-b}-\frac{2a}{v}-\tau\ln n_{_Q},
\end{eqnarray}
where $n_{_Q}$ is ``quantum concentration''. The quantum concentration $n_{_Q}$, as pioneered by Kittel and Kroemer~\cite{Kittel1980}, is the concentration (i.e. the number of particles per unit volume) of a system where the interparticle distance is roughly equal to the thermal average de Broglie wavelength. It has made quantum states ``countable'' and therefore accessible for a statistical calculation of entropies. This concentration will keep turning up in the thermal physics of gases, in semi-conductor theory and in the theory of chemical reactions. It is defined as
 \begin{eqnarray}
n_{_Q}=\left(\frac{m\tau}{2\pi \hbar^2}\right)^{3/2},
\end{eqnarray}
where $m$ is the mass of the particles (like molecule or atom) in the system and $\hbar$ is reduced Planck's constant.

The particle concentration is $n=1/V$. Whenever $n/n_{_Q}\ll 1$, we say that the gas is in the classical regime. An ideal gas is defined as a gas of noninteracting atoms in the classical regime. Quantum effects become appreciable when the particle concentration is greater than or equal to the quantum concentration.

\begin{figure}[htb]
\begin{center}
\subfigure[$\tau=0.900\tau_c$]{\includegraphics[width=55mm]{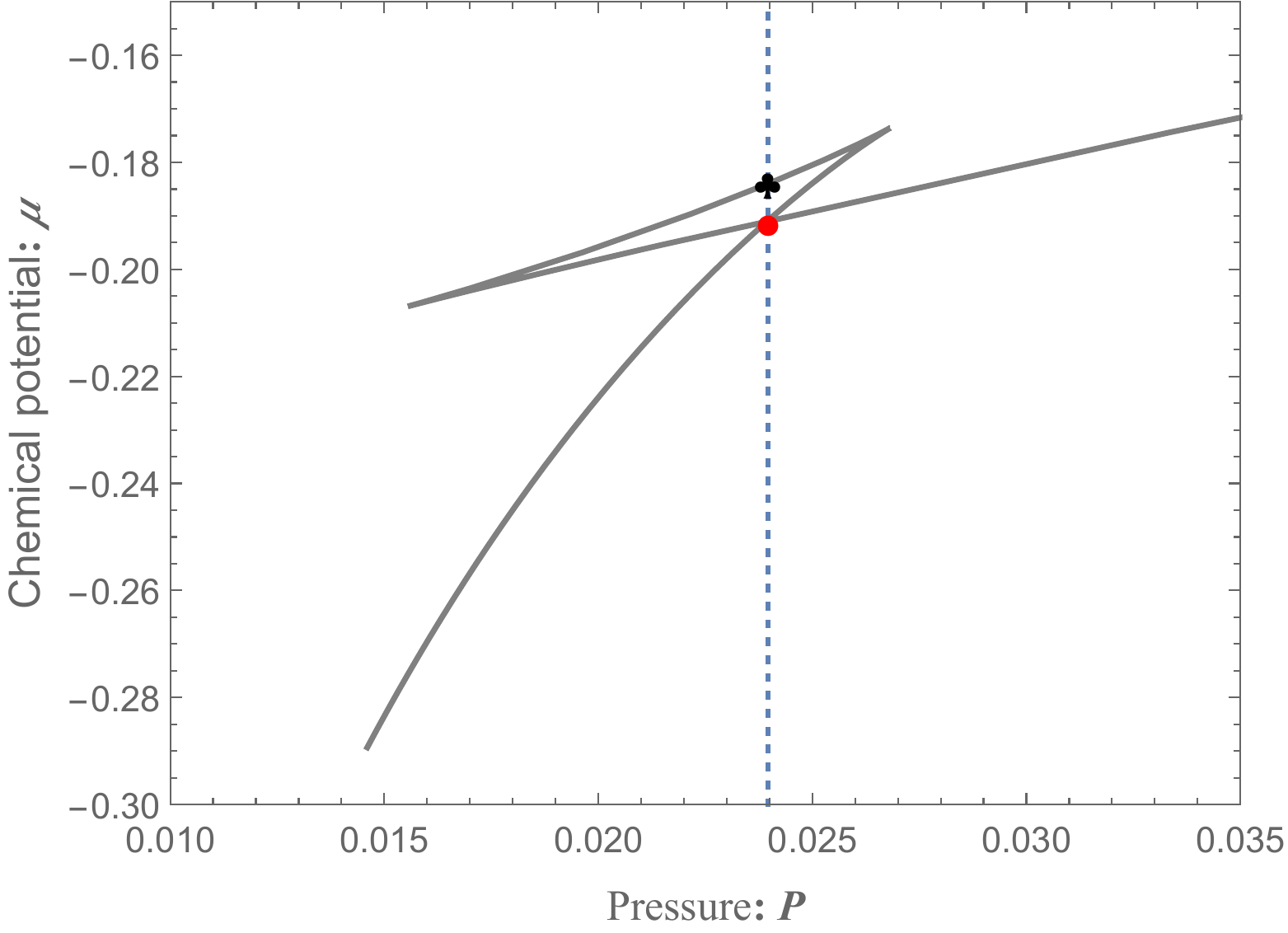}}
\qquad
\subfigure[$\tau=0.900\tau_c$ and $P=0.647P_c$]{\includegraphics[width=55mm]{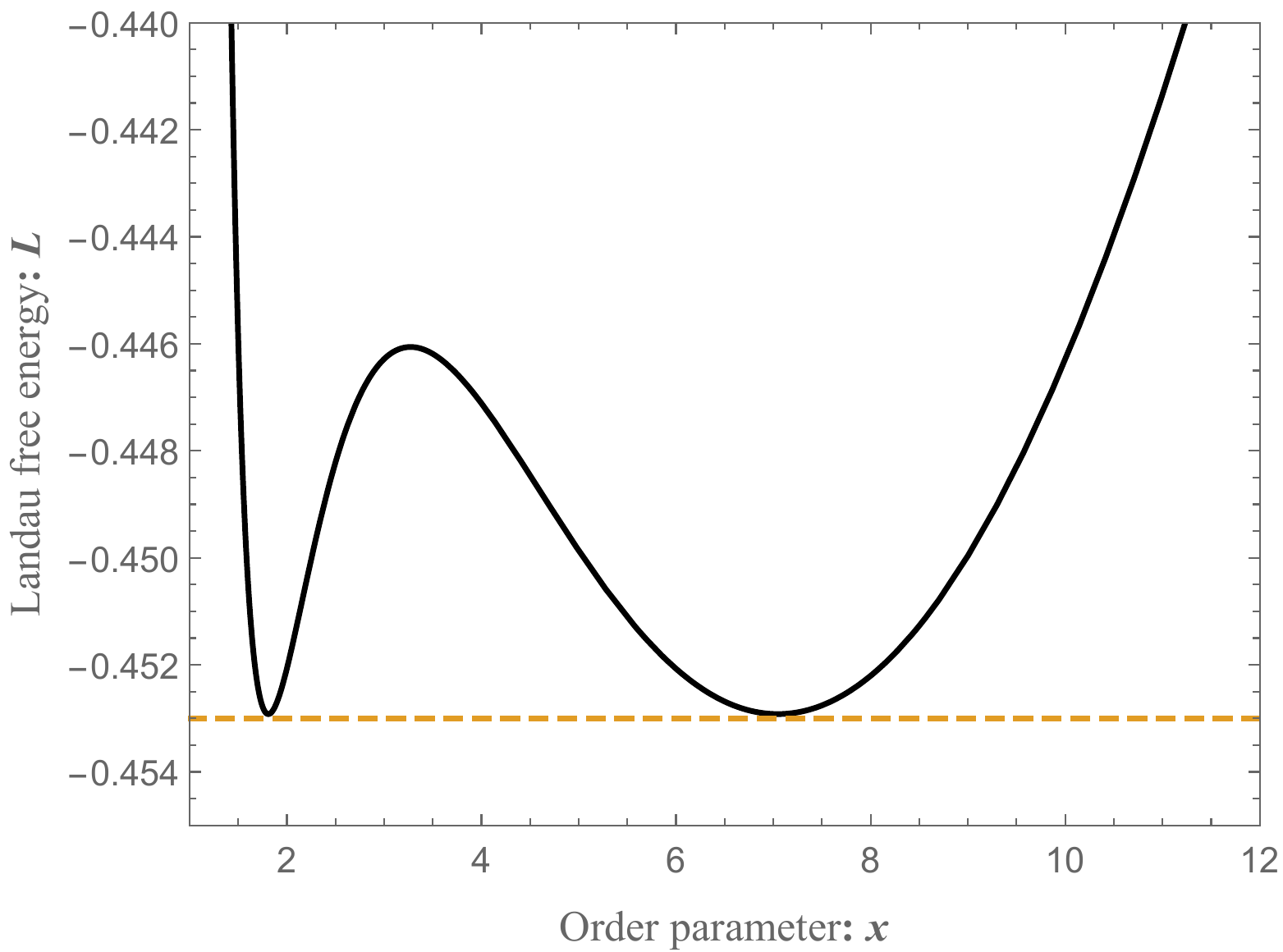}}
\end{center}
\caption{(color online) The behaviors of the chemical potential $\mu$ with respect to the pressure $P$ and the Landau functional $L$ with respect to the order parameter $x$ for the van der Waals fluid, where $\tau_c=8a/(27b)$ and $P_c=a/(27b^2)$. The red point \red{$\bullet$} in diagram (a) corresponds to two equal global minimum points in diagram (b), and the black point {\scriptsize$\clubsuit$} corresponds to the local maximum point.}
\label{fig0}
\end{figure}

According to Eqs.~(\ref{landau}),~(\ref{landauf}) and~(\ref{eosvdw}), we can obtain the expression of the Landau functional for the van der Waals fluid, which is similar to that in~\cite{Huang1987}, but is different in form,
\begin{eqnarray}\label{landauvdw}
L=Px-\frac{a}{x}-\tau\ln(x-b),
\end{eqnarray}
where $x$ is regarded as the order parameter. Compared Eqs.~(\ref{chemical}) and~(\ref{landauvdw}), it can be clearly observed that Landau free energy has the dimension of energy, but it is different from Gibbs free energy or chemical potentials. We can verify that the position of the minimum of the Landau functional exactly corresponds to the real state of the van der Waals fluid system,
\begin{eqnarray}
\frac{\text{d}L}{\text{d}x}=0 \qquad \Rightarrow \qquad P=\frac{\tau}{v-b}-\frac{a}{v^2}.
\end{eqnarray}
In order to be more intuitive, we give the behavior of the chemical potential of the van der Waals system with respect to the pressure and Landau functional with respect to the order parameter in Fig.~\ref{fig0}. It can be seen that the behaviors of Gibbs free energy and Landau functional have an exact corresponding relationship. The swallow tail intersection of Gibbs free energy corresponds to two minimum points of Landau functional, which means that the system is in a stable coexistence state of gas and liquid. If we adjust the pressure of the system, that is, the position of the dotted line in diagram (a), we can directly understand that the extreme point in diagram (b) will move up or down, which implies that the system will be more inclined to be in a single-phase stable state. The behavior of Landau functional clearly shows us the phase structure of the van der Waals system.

\section{For RN-AdS black hole}
For a four-dimensional charged AdS black hole, i.e., Reissner-Nordstr\"{o}m AdS black hole, its metric is~\cite{Kubiznak2012}
\begin{eqnarray}
\text{d}s^2=-f(r)\text{d}t^2+\frac{\text{d}r^2}{f(r)}+r^2 (\text{d}\theta^2+\sin^2 \theta \text{d}\varphi^2),
\end{eqnarray}
where the function $f(r)$ is
\begin{equation*}
f(r)=1-\frac{2M}{r}+\frac{r^2}{l^2}+\frac{Q^2}{r^2},
\end{equation*}
and $M$ represents the ADM mass of the black hole (or the thermodynamic enthalpy in the extended phase space), $Q$ is the total charge of the black hole and $l$ is the AdS radius which can be related to the thermodynamic pressure via~\cite{Kastor2009}
\begin{eqnarray}\label{pressure}
P=\frac{3}{8\pi l^2}.
\end{eqnarray}
In addition, we can also get the relationship between the mass or enthalpy of the black hole and the radius of the event horizon $r_h$ satisfied equation $f(r_h)=0$,
\begin{eqnarray}\label{enthalpy}
M=\frac{r_h}{2}+\frac{4\pi P r_h^3}{3}+\frac{Q^2}{2r_h^2}.
\end{eqnarray}
Hence, the thermodynamic volume conjugated with pressure is
\begin{eqnarray}\label{volume}
V=\frac43\pi r_h^3=\beta^3r_h^3.
\end{eqnarray}
Based on the `Euclidean trick', we can identify the temperature of the charged AdS black hole is~\cite{Kubiznak2012}
\begin{eqnarray}\label{temperature}
T=\frac{1}{4\pi r_h}\left(1+8\pi P r_h^2-\frac{Q^2}{r_h^2}\right),
\end{eqnarray}
Meanwhile, the entropy which conjugated with the temperature is
\begin{eqnarray}\label{entropy}
S=\pi r_h^2.
\end{eqnarray}

We have known that the thermodynamic critical behavior of the charged AdS black hole is very consistent with that of van der Waals fluid, which will undergo gas-liquid phase transition. It is called the large-small black hole phase transition for the charged Ads black hole. For the black hole system, its critical values are~\cite{Kubiznak2012}
\begin{eqnarray}\label{criticals}
r_c=\sqrt{6}Q, \quad V_c=6\beta^3\sqrt{6}Q^3, \quad T_c=\frac{\sqrt{6}}{18\pi Q}, \quad P_c=\frac{1}{96\pi Q^2}.
\end{eqnarray}

\begin{figure}[htb]
\begin{center}
\subfigure[At $p<p_c$.]{\includegraphics[width=55mm]{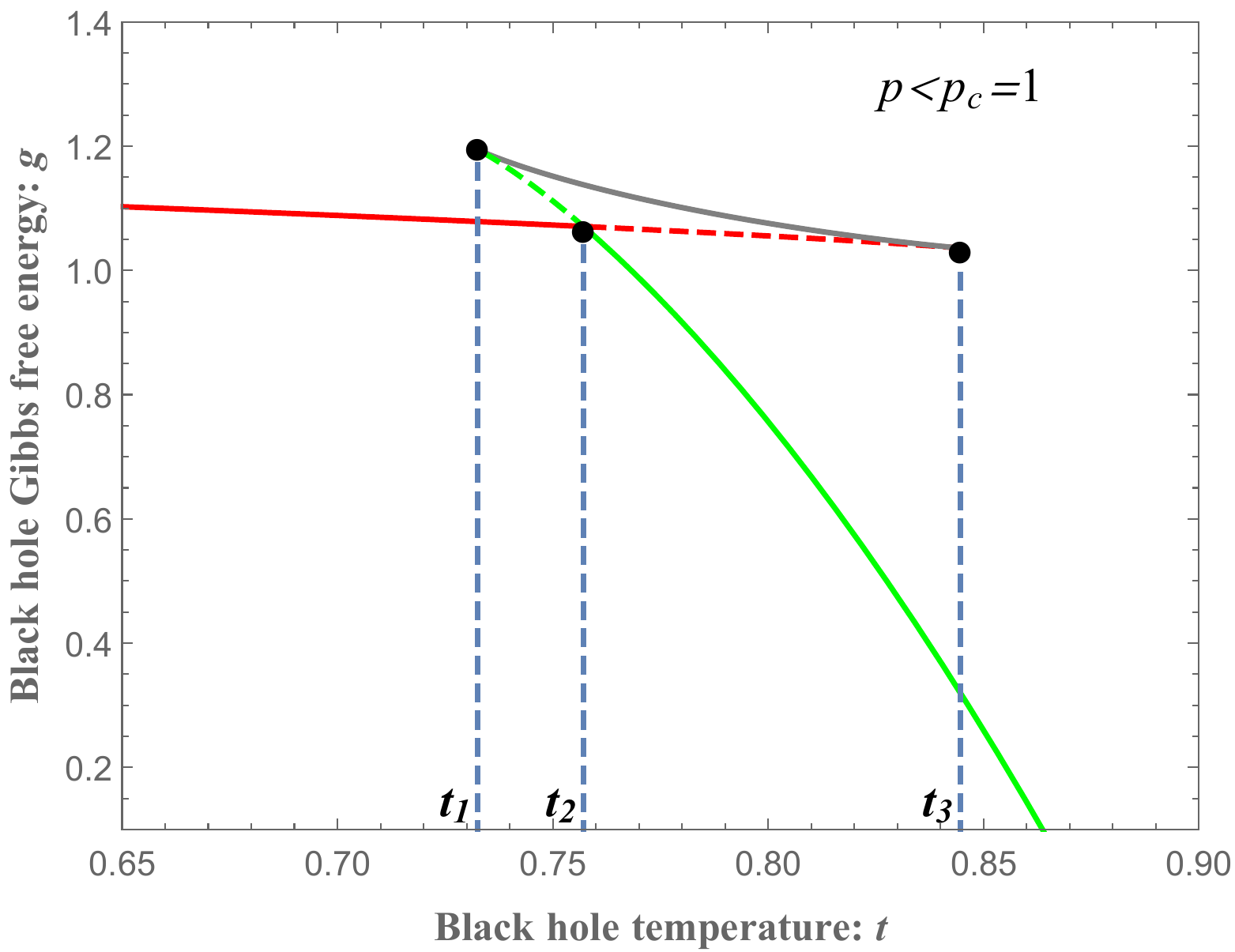}}
\qquad
\subfigure[At the different pressure.]{\includegraphics[width=55mm]{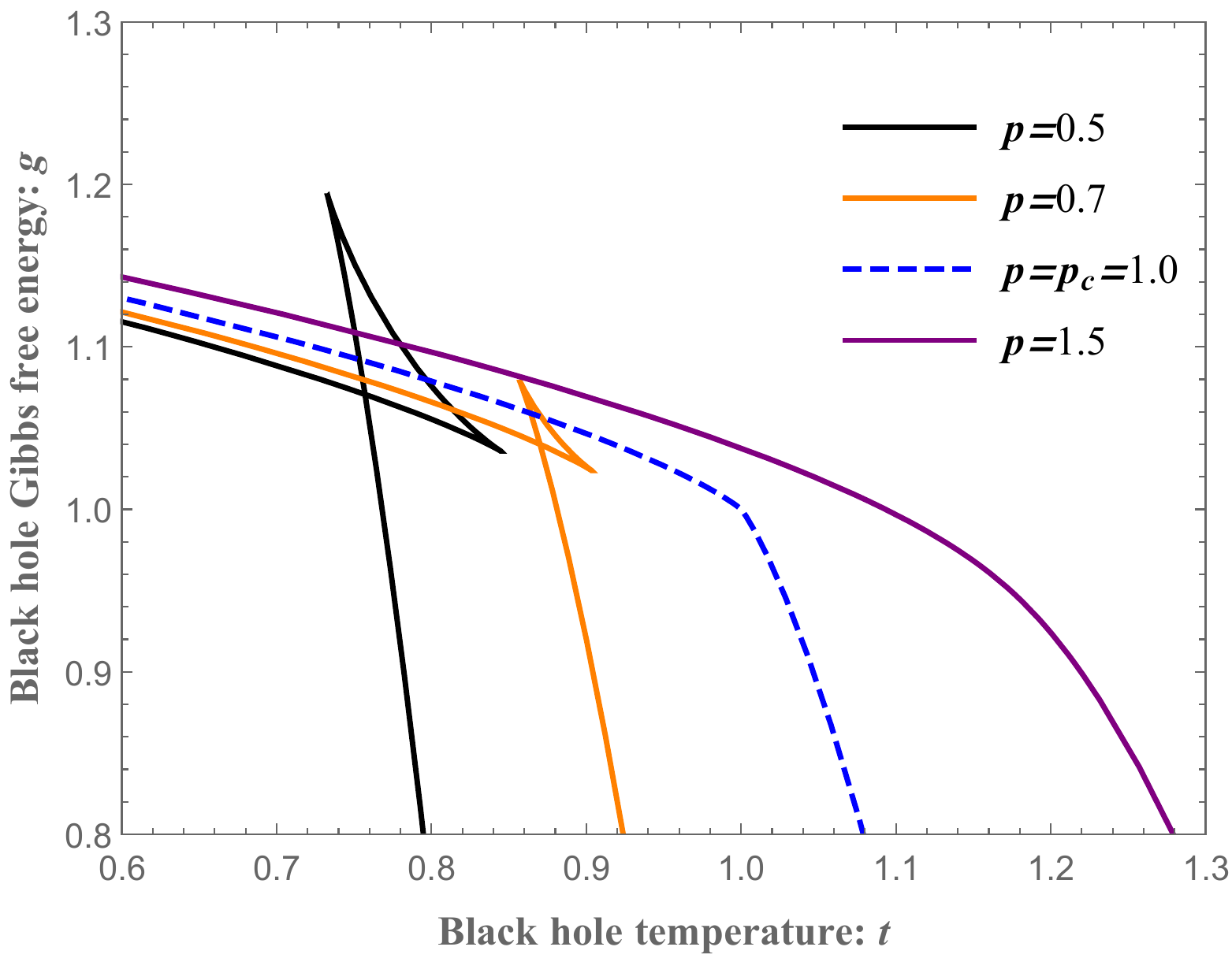}}
\end{center}
\caption{(color online) The behaviors of the dimensionless Gibbs free energy $g$ with respect to the dimensionless temperature $t$ at the different dimensionless pressure $p$ for the charged AdS black hole, respectively.}
\label{fig1}
\end{figure}

Gibbs free energy plays a key role in the analysis of thermodynamic phase transition behavior. For the charged AdS black hole, its expression is in terms of Eqs.~(\ref{enthalpy}),~(\ref{temperature}) and~(\ref{entropy})
\begin{eqnarray}
G\equiv M-T S=\frac{1}{4}\left(r_h-\frac{8\pi}{3}P r_h^3+\frac{3Q^2}{r_h}\right).
\end{eqnarray}
Of course, at the critical point~(\ref{criticals}), the Gibbs free energy becomes $G_c=\sqrt{6}Q/3$. For simplicity, several dimensionless parameters are introduced
\begin{eqnarray}\label{reparameters}
t:=\frac{T}{T_c}, \quad p:=\frac{P}{P_c}, \quad z:=\frac{r_h}{r_c}, \quad g:=\frac{G}{G_c}.
\end{eqnarray}
Hence the temperature and Gibbs free energy of the charged AdS black hole can be rewritten as~\cite{Kubiznak2012,Spallucci2013}
\begin{eqnarray}
t&=&\frac{3}{4}\left(\frac{1}{z}+\frac{pz}{2}-\frac{1}{6z^3}\right),\\
g&=&\frac{3}{4}\left(z-\frac{pz^3}{6}+\frac{1}{2z}\right).
\end{eqnarray}
For different pressures, the Gibbs free energy versus temperature is shown in Fig.~\ref{fig1}. A stable thermodynamic system always tends to the state with lower Gibbs free energy. We can clearly observe from Fig.~\ref{fig1} that when the pressure of the black hole system is lower than the critical pressure $p_c$. The stable black hole state is described by the red solid line and green solid line. When the pressure tends to the critical pressure, the swallow tail structure disappears, and the black hole will have a second-order phase transition or continuous phase transition. Once the pressure exceeds the critical pressure, the whole system is always in a stable state. In the critical region, which is the typical representative in diagram (a) in Fig.~\ref{fig1}, we can see that there are three key temperature values.
\begin{itemize}
  \item The temperature of swallow tail intersection $t_2$ can be calculated by Maxwell equal area law and its value is~\cite{Spallucci2013}
  \begin{eqnarray}\label{t2}
   t_2=\sqrt{\frac{p(3-\sqrt{p})}{2}}.
  \end{eqnarray}
  \item The temperatures of swallow tail tip $t_1$ and $t_3$ are local minimum and local maximum satisfied equation $\partial t/\partial z=0$. Thus we have
  \begin{eqnarray}\label{t13}
   t_1=\frac{3-p+3\sqrt{1-p}}{2p\left(\frac{\sqrt{1-p}+1}{p}\right)^{3/2}}, \qquad
   t_3=\frac{ \left(2-\sqrt{1-p}\right)\sqrt{1+\sqrt{1-p}}}{2}.
  \end{eqnarray}
\end{itemize}

Next, we investigate the behavior of the newly introduced Landau functional. According to Eq.~(\ref{temperature}), we can read $F(X,T,P)=P-\beta T/(2X^{1/3})+\beta^2/(8\pi X^{2/3})-\beta^4 Q^2/(8\pi X^{4/3})$.
Hence the Landau functional of the charged AdS black hole can be written as
\begin{eqnarray}
L=PX-\frac{3\beta T}{4}X^{2/3}+\frac{3\beta^2}{8\pi}X^{1/3}+\frac{3\beta^4 Q^2}{8\pi}X^{-1/3}.
\end{eqnarray}
When $X=V$ and taking Eq.~(\ref{temperature}) into the above expression, we have $L=G$, which implies that the Landau functional $L$ has dimension of energy, and is related to, but is not identical with the Gibbs free energy of the system. For the sake of simplicity, with the help of~(\ref{reparameters}), we can introduce the dimensionless Landau functional $\psi(x,t,p)$
\begin{eqnarray}
\psi(x,t,p):=\frac{L}{G_c}=\frac{1}{4}\left(\frac{1}{x}+6x+px^3-4t x^2\right), \qquad x:=\left(\frac{X}{V_c}\right)^{1/3}.
\end{eqnarray}

\begin{figure}[htb]
\begin{center}
\includegraphics[width=60mm]{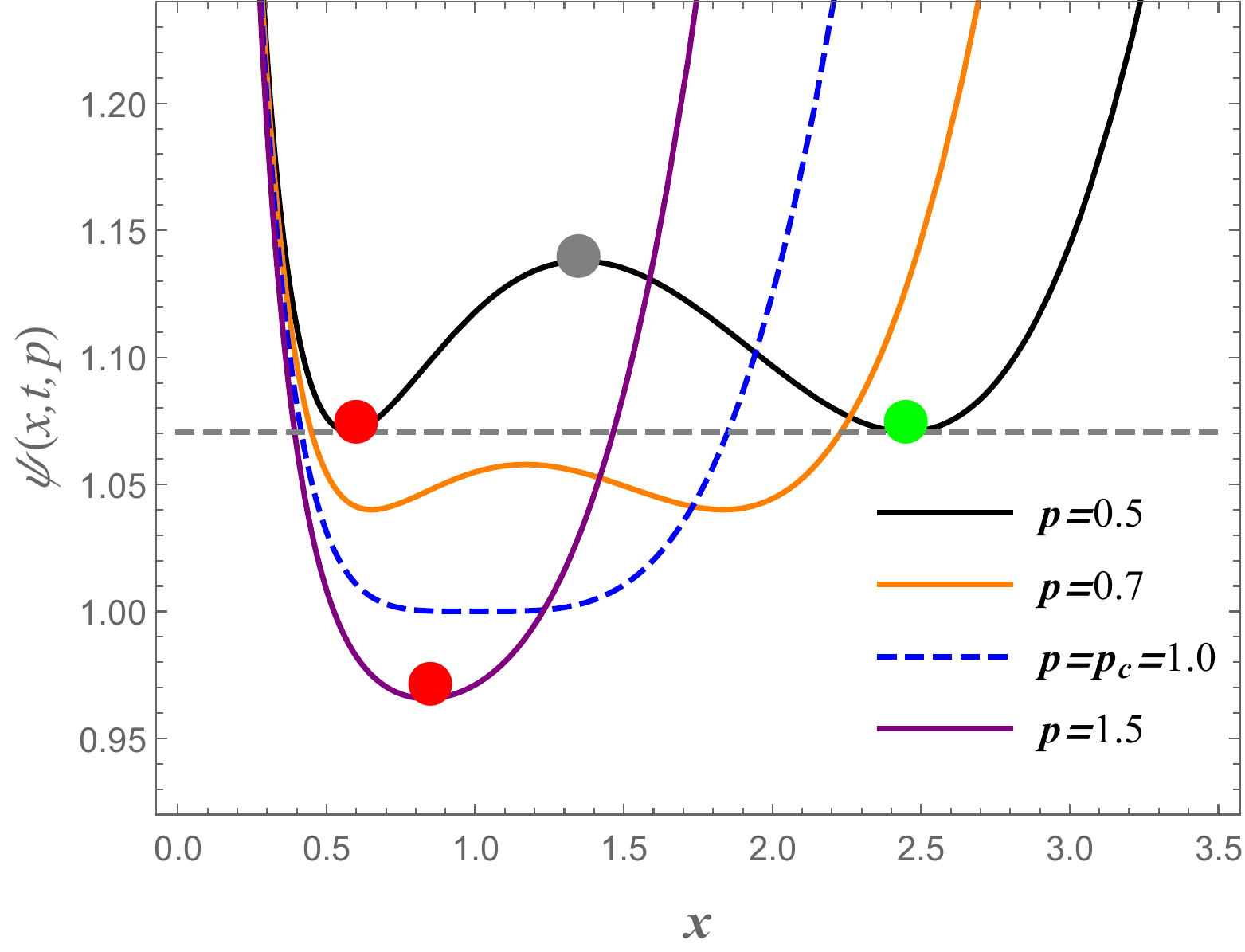}
\end{center}
\caption{(color online) Landau functional $\psi(x,t,p)$ versus the order parameter $x$ at $t=t_2$ and different $p$ respectively.}
\label{fig2}
\end{figure}

\begin{figure}[htb]
\begin{center}
\subfigure[]{\includegraphics[width=50mm]{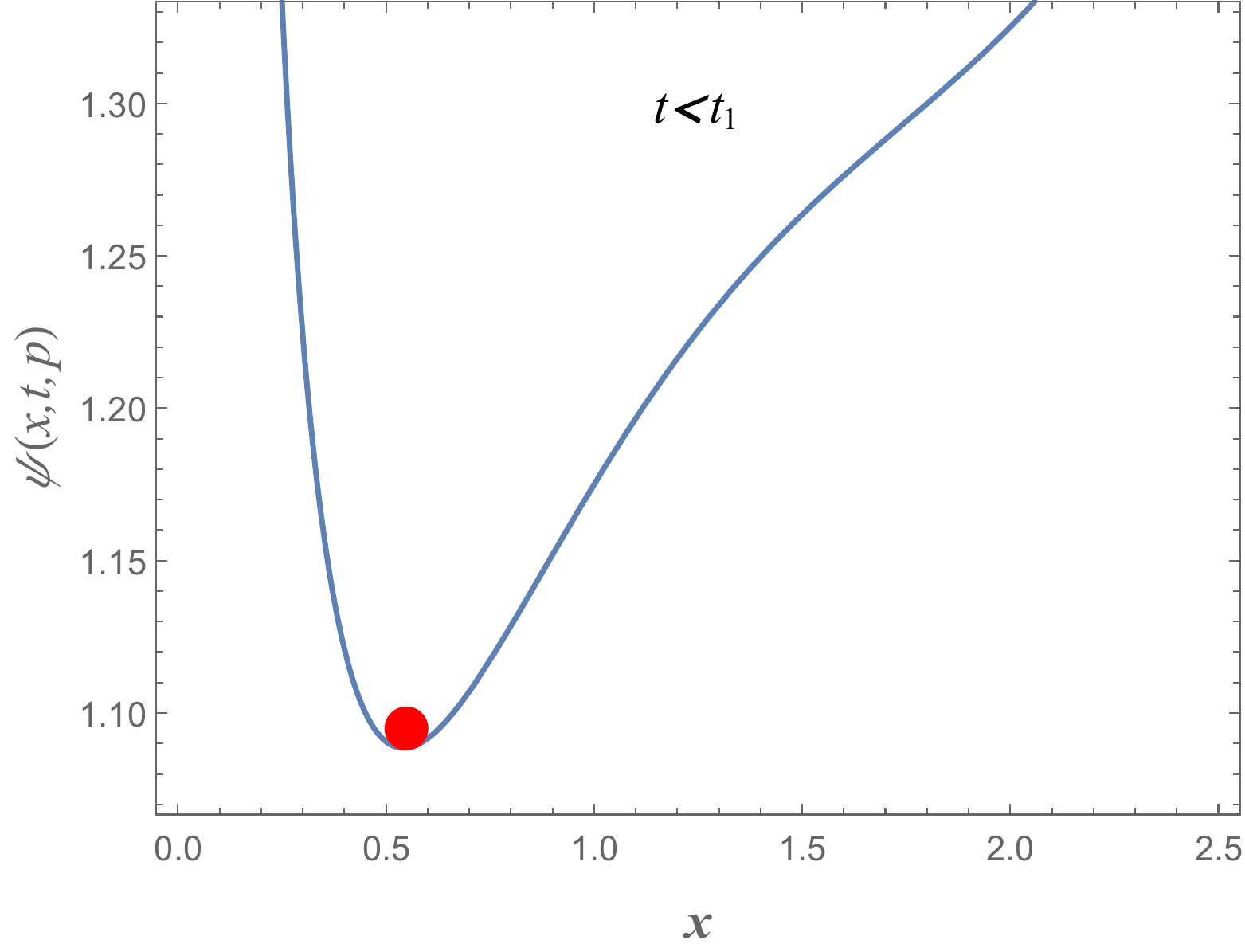}}
\quad
\subfigure[]{\includegraphics[width=50mm]{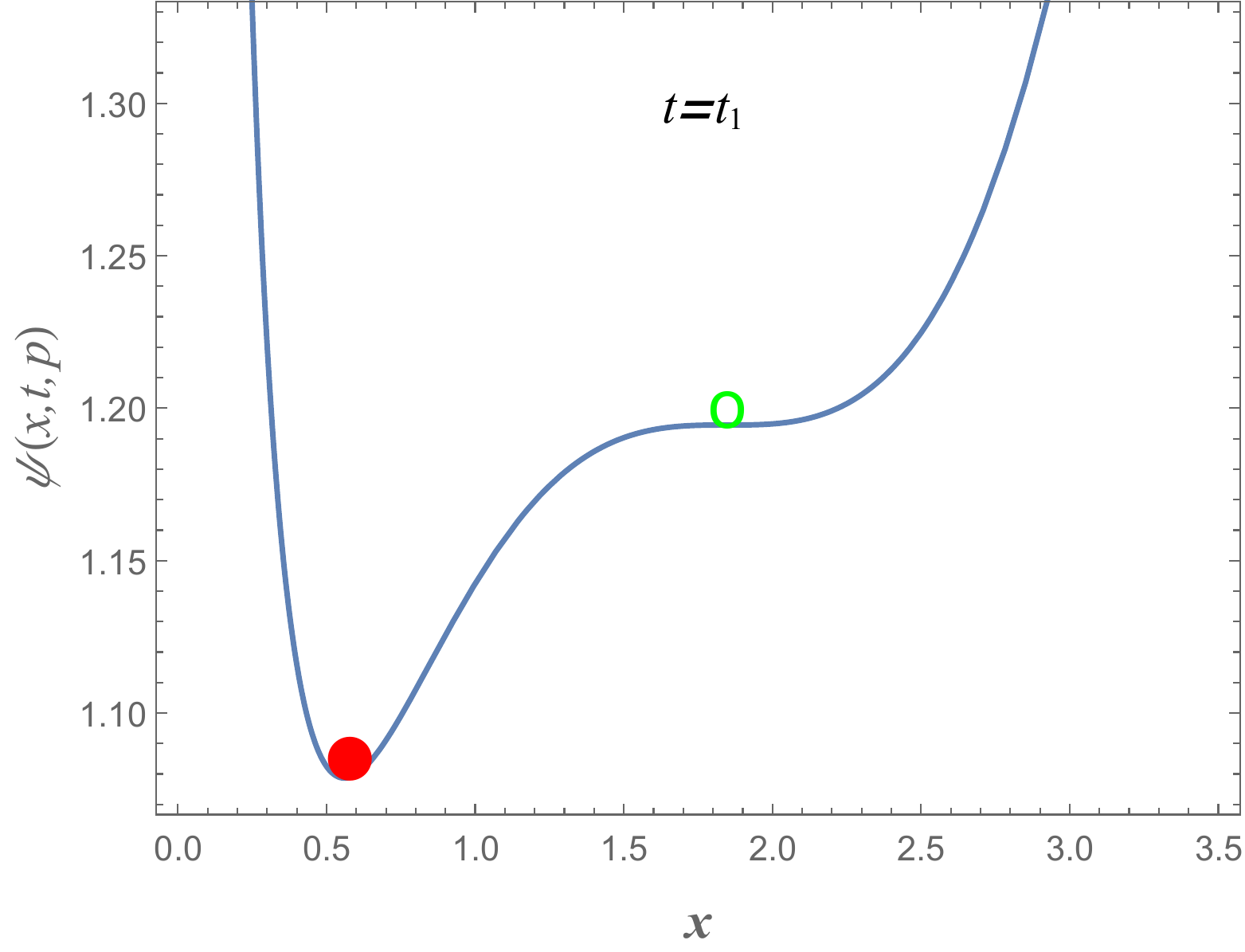}}
\quad
\subfigure[]{\includegraphics[width=50mm]{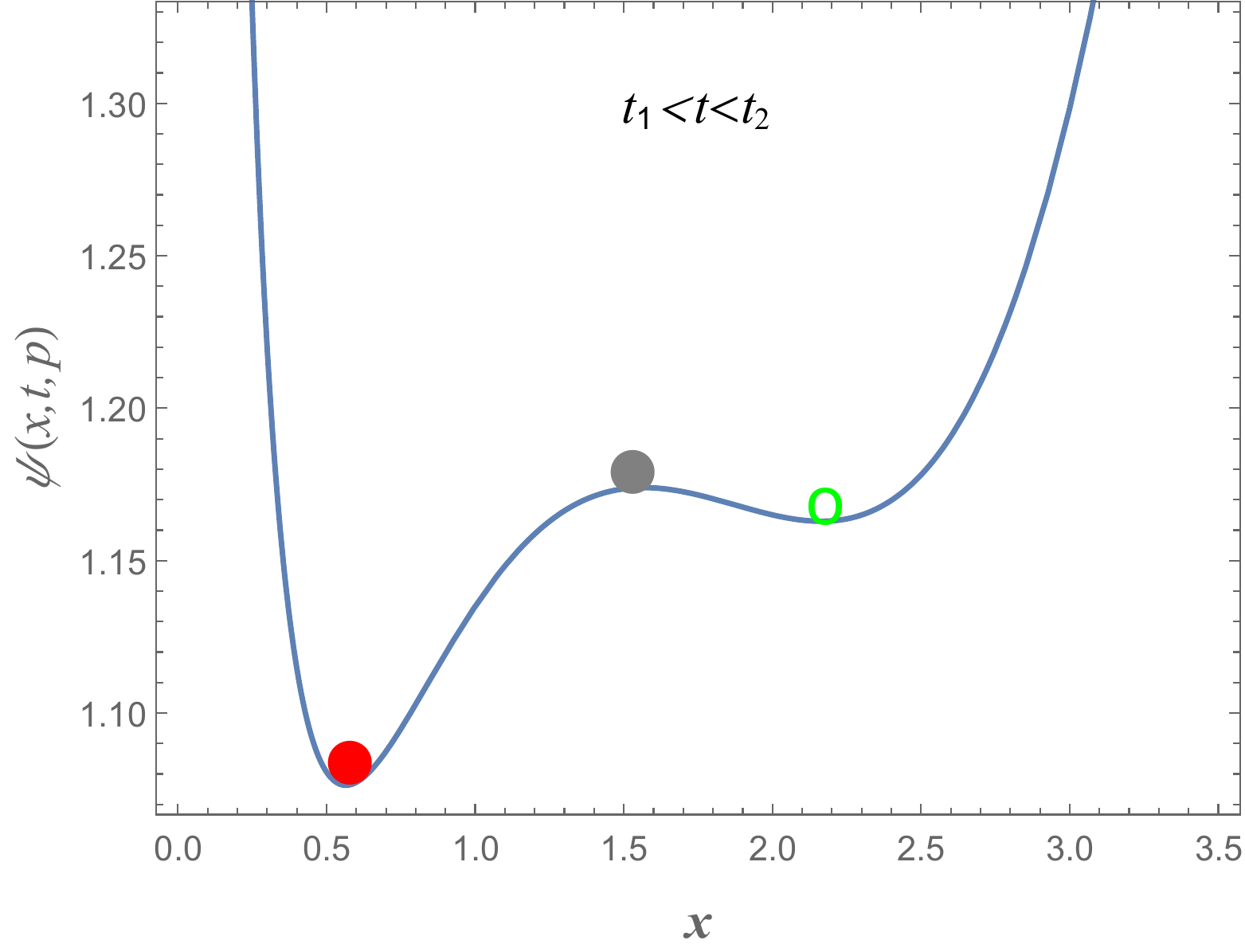}}\\
\subfigure[]{\includegraphics[width=50mm]{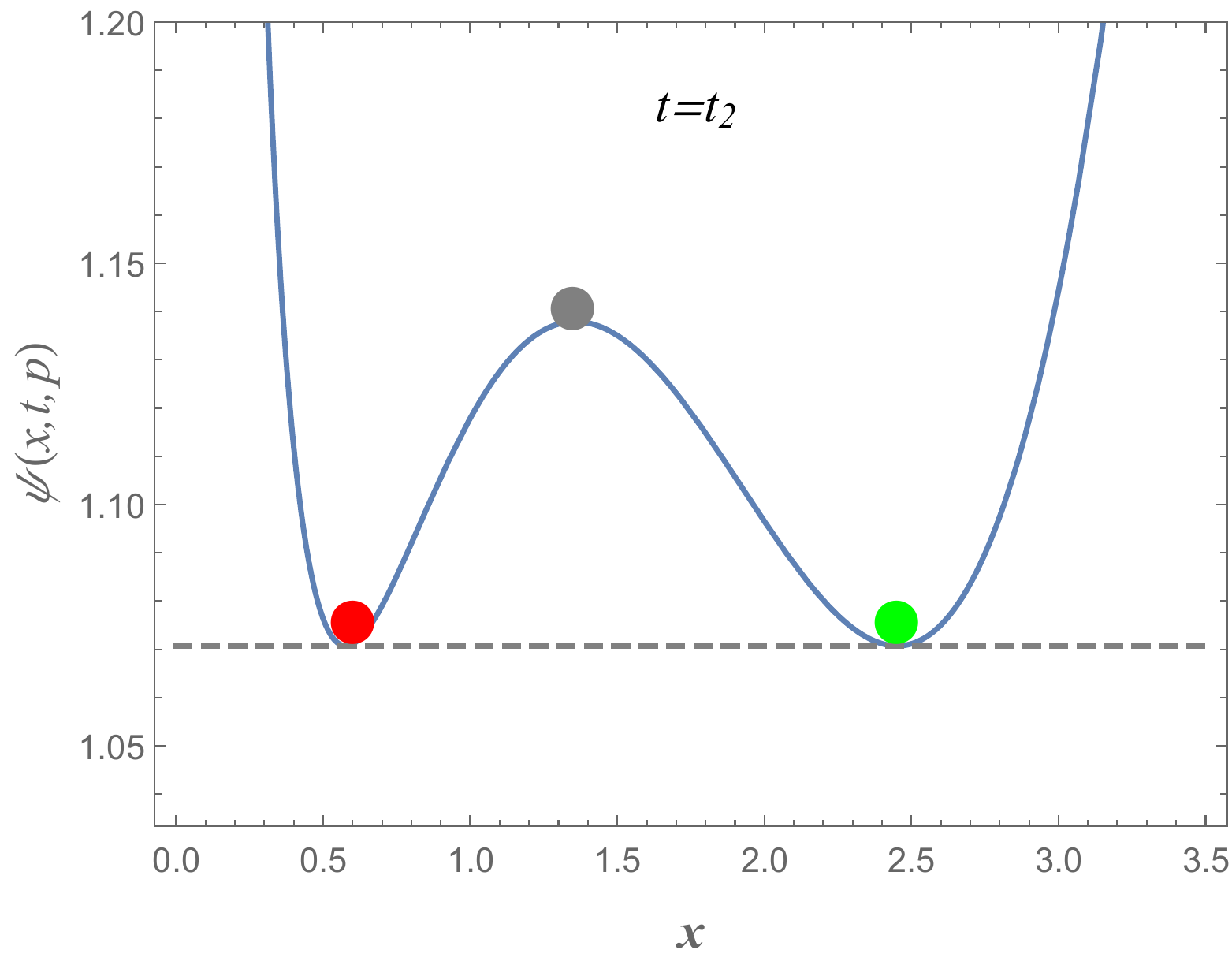}}\\
\subfigure[]{\includegraphics[width=50mm]{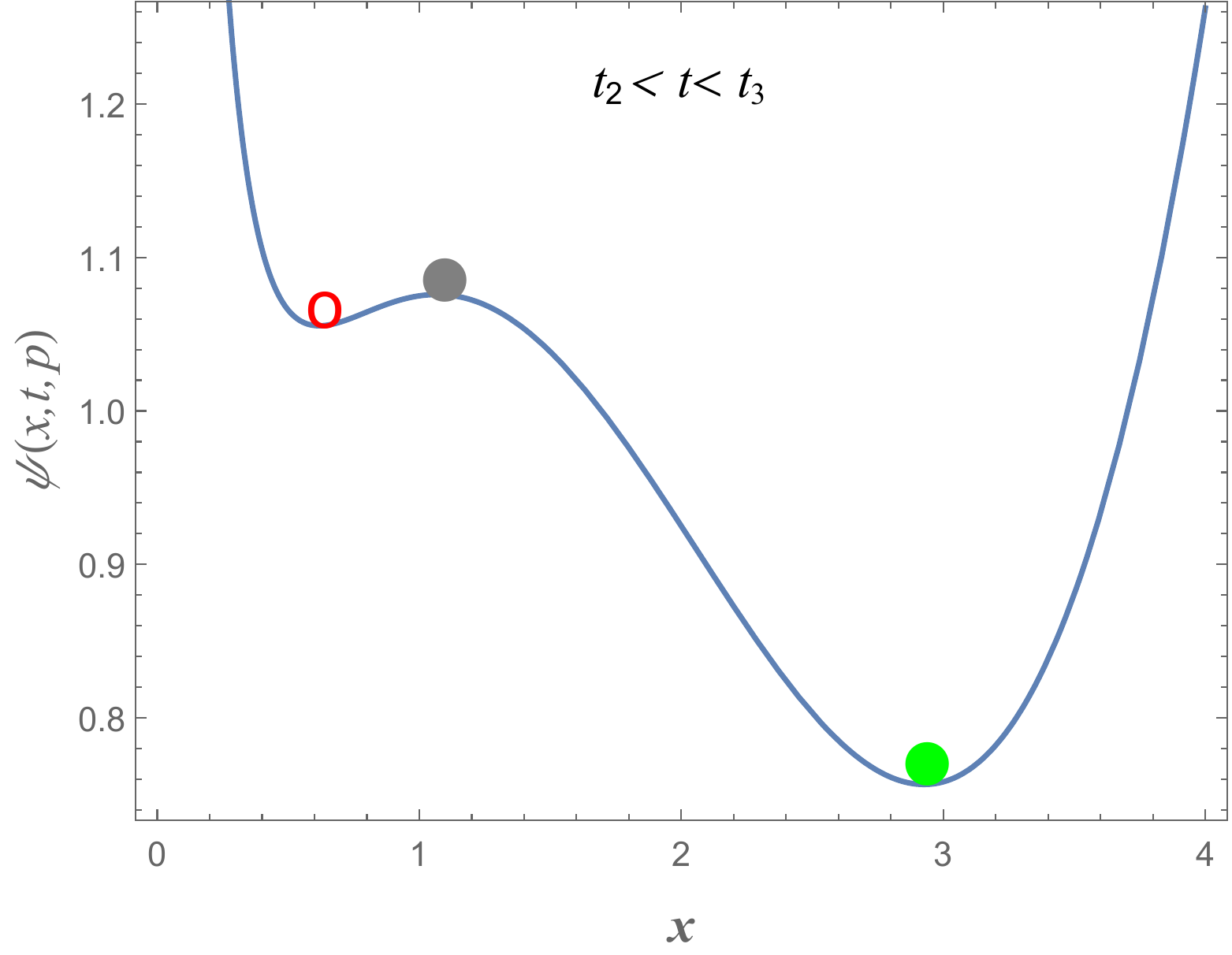}}
\quad
\subfigure[]{\includegraphics[width=50mm]{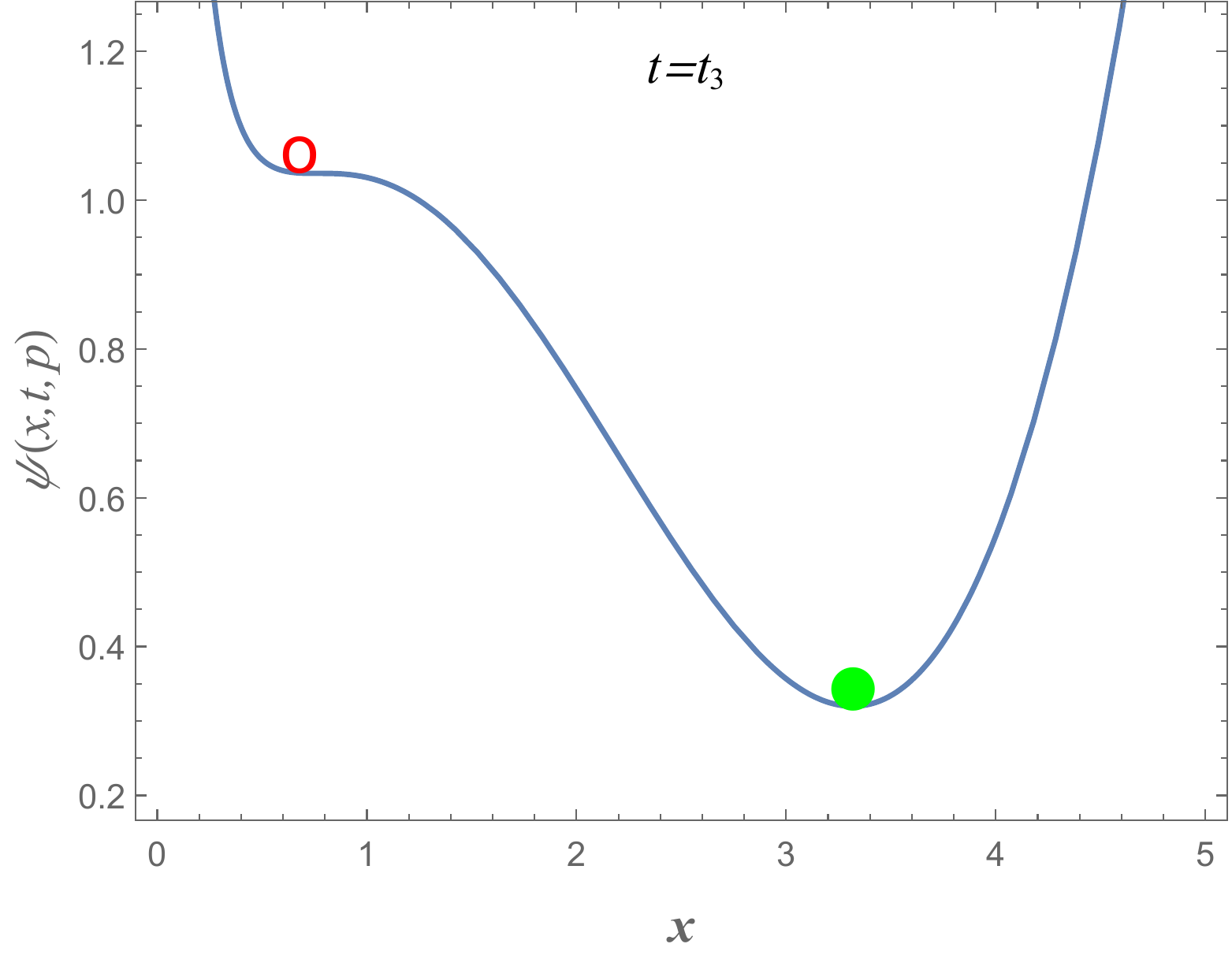}}
\quad
\subfigure[]{\includegraphics[width=50mm]{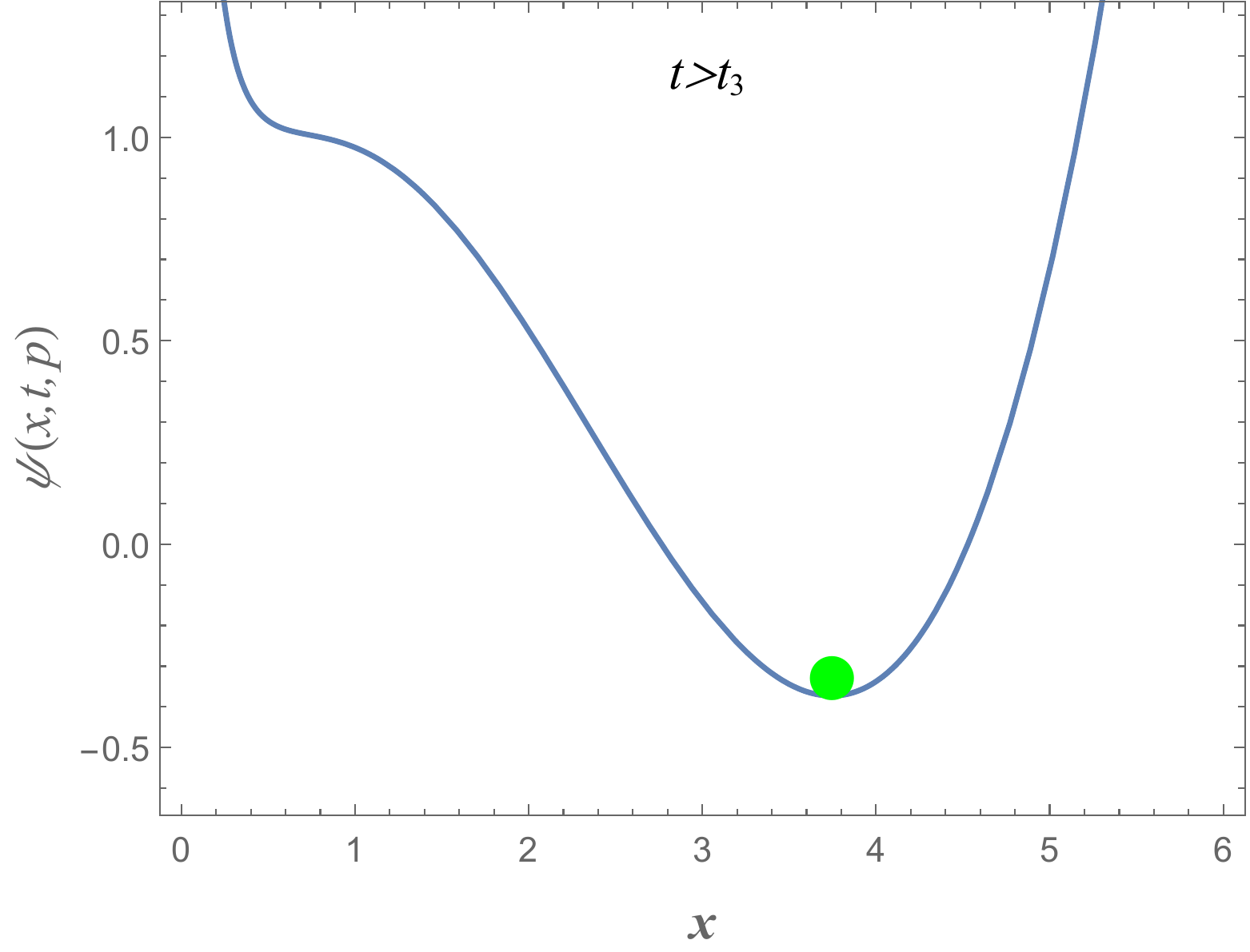}}
\end{center}
\caption{(color online) Landau functional $\psi(x,t,p)$ versus the order parameter $x$ at $p=0.5$ and different $t$  respectively.}
\label{fig3}
\end{figure}

We have drawn the relationship between Landau functional $\psi(x,t,p)$ and the order parameter $x$ at different temperature $t$ and pressure $p$ in Figs.~\ref{fig2} and~\ref{fig3}, from which we can clearly see the phase transition process of the charged AdS black hole. The specific descriptions are as follows.
\begin{itemize}
  \item From Fig.~\ref{fig2}, when $t=t_2$, with the decreasing of pressure $p$, the global minima of free energy splits into two equivalent global minima. Meanwhile a local maximum of free energy appears. Physically, at higher pressure, there is only one global minimum of the free energy, which corresponds to a phase of the black hole, labeled as \red{\Large$\bullet$}-phase (small black hole phase). As the pressure decreases and passes through the critical pressure $p_c=1$, the two equivalent global minima of free energy correspond to \red{\Large$\bullet$}-phase and \green{\Large$\bullet$}-phase (large black hole phase) respectively, while the new local maximum of free energy corresponds to \textcolor[rgb]{0.50,0.50,0.50}{\Large$\bullet$}-phase (middle black hole phase). Obviously, \red{\Large$\bullet$}-phase and \green{\Large$\bullet$}-phase are stable, while \textcolor[rgb]{0.50,0.50,0.50}{\Large$\bullet$}-phase is unstable which can be regarded as an intermediate transition phase. Because the two global minima of free energy are equal, it means that the chances of \red{\Large$\bullet$}-phase going through \textcolor[rgb]{0.50,0.50,0.50}{\Large$\bullet$}-phase to \green{\Large$\bullet$}-phase and \green{\Large$\bullet$}-phase going through \textcolor[rgb]{0.50,0.50,0.50}{\Large$\bullet$}-phase to \red{\Large$\bullet$}-phase are the same. \red{\Large$\bullet$}-phase and \green{\Large$\bullet$}-phase are in two-phase coexistence. At this time, the value of pressure determines the degree of difficulty when \red{\Large$\bullet$}-phase or \green{\Large$\bullet$}-phase reaches \textcolor[rgb]{0.50,0.50,0.50}{\Large$\bullet$}-phase. The smaller the pressure is, the harder it is to reach, and vice versa. The phase transition described in these three figures is called continuous phase transition, or second-order phase transition.
  \item From figures (a) to (d) in Fig.~\ref{fig3} in reverse order, at fixed pressure $p<p_c$, as the temperature $t$ decreases from $t_2$, it can be clearly seen that two equivalent global minima of free energy begin to change, i.e., one is still a global minimum while the other becomes a local minimum until it disappears. Specifically, when the temperature $t$ decreases from $t_2$ to $t_1$, the free energy of \red{\Large$\bullet$}-phase is lower than that of \green{\Large$\bullet$}-phase, which means that the black hole system tends to be in \red{\Large$\bullet$}-phase. In this process, \green{\Large$\bullet$}-phase changes into \red{\Large$\bullet$}-phase through \textcolor[rgb]{0.50,0.50,0.50}{\Large$\bullet$}-phase. Naturally, \red{\Large$\bullet$}-phase is still stable while \green{\Large$\bullet$}-phase becomes metastable (marked as \green{\Large $\circ$}-phase). Once the temperature $t$ is lower than $t_1$, the whole black hole system will be in \red{\Large$\bullet$}-phase completely. From these four figures, we can clearly see that the global minimum of free energy has changed, which is called the first-order phase transition, that is, the first-order phase transition from \green{\Large$\bullet$}-phase to \red{\Large$\bullet$}-phase.
  \item From figures (d) to (g) in Fig.~\ref{fig3}, at fixed pressure $p<p_c$, as the temperature $t$ increases from $t_2$, it can be also clearly seen that two equivalent global minima of free energy begin to change. Specifically, when the temperature $t$ increases from $t_2$ to $t_3$, the free energy of \green{\Large$\bullet$}-phase is lower than that of \red{\Large$\bullet$}-phase, which means that the black hole system tends to be in \green{\Large$\bullet$}-phase. In this process, \red{\Large$\bullet$}-phase changes into \green{\Large$\bullet$}-phase through \textcolor[rgb]{0.50,0.50,0.50}{\Large$\bullet$}-phase. Moreover, \green{\Large$\bullet$}-phase is still stable while \red{\Large$\bullet$}-phase becomes metastable (marked as \red{\Large $\circ$}-phase). Once the temperature $t$ is higher than $t_3$, the whole black hole system will be in \green{\Large$\bullet$}-phase completely. From these figures, we can clearly see the first-order phase transition from \red{\Large$\bullet$}-phase to \green{\Large$\bullet$}-phase.
\end{itemize}

From the analysis of the Landau functional, we can clearly see a more visual process of black hole phase transition. The splitting of the global minimum of Landau functional reflects the continuous phase transition (second-order phase transition) of the black hole, and the transformation between the global minimum of free energy reflects the first-order phase transition of the black hole. These analyses are consistent with that of the Gibbs free energy, but here we can clearly and intuitively see that how the transition between the different phases of the black hole occurs. When the temperature parameter is fixed, the pressure parameter determines the occurrence of continuous phase transition, while when the pressure parameter is fixed, the change of temperature parameter affects the behavior of first-order phase transition. Here, three typical temperature parameters ($t_1, t_2, t_3$) play a key role, which is actually a constraint relationship between pressure parameter and temperature parameter, see Eqs.~(\ref{t2}) and~(\ref{t13}). Different constraints have different effects on the phase transition behavior of black holes.

\begin{figure}[htb]
\begin{center}
\includegraphics[width=140mm]{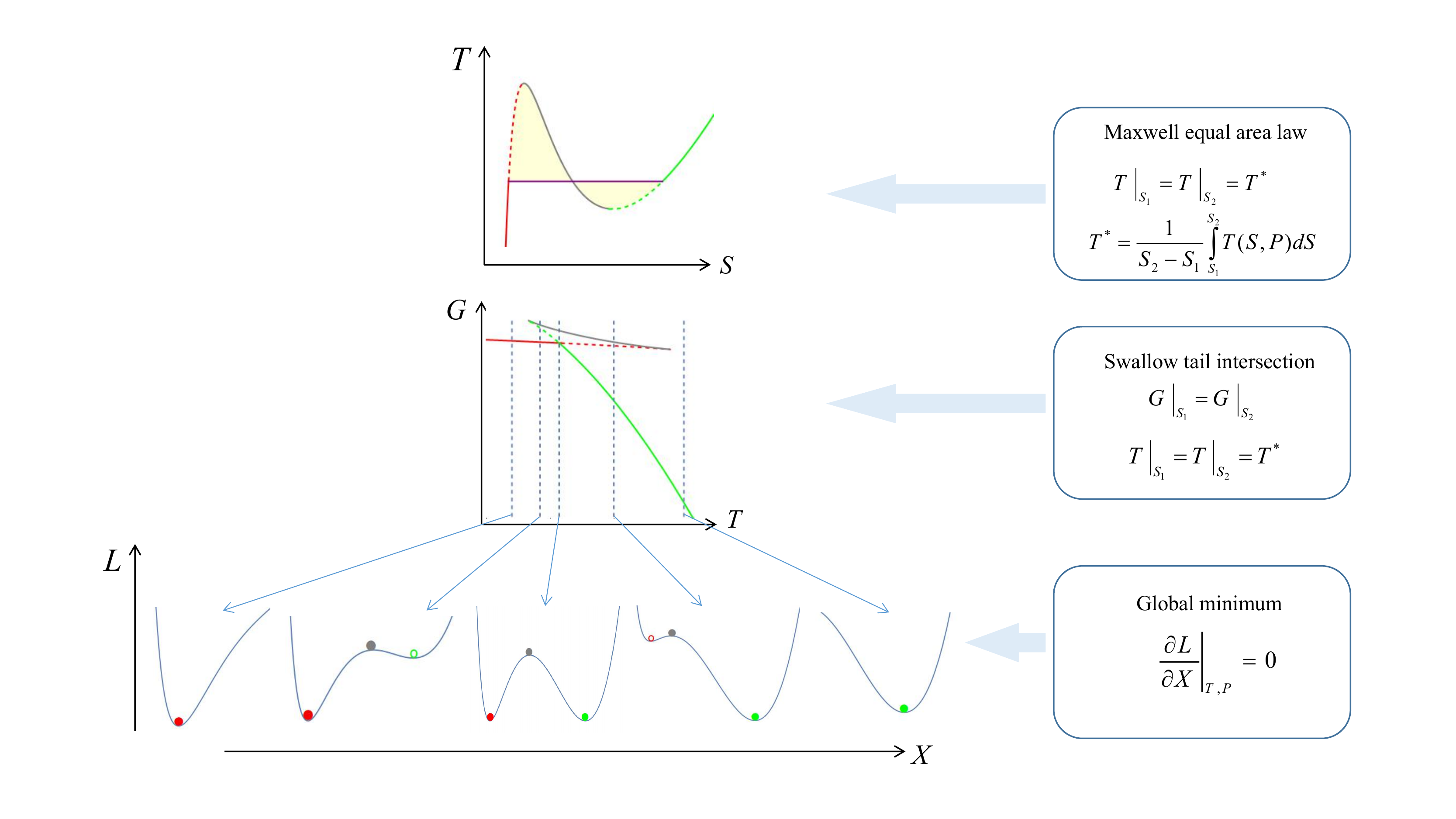}
\end{center}
\caption{(color online) Three equivalence schemes of black hole phase transition at fixed pressure $P<P_c$: Maxwell equal area law in the equation of state, Gibbs free energy and Landau functional. Curves and spheres of the same color (red, green and gray) represent the same phases, the solid line corresponds to the solid sphere, and the dotted line corresponds to the hollow sphere.}
\label{fig4}
\end{figure}

\section{Summary}
In this paper, we construct the Landau free energy or Landau functional in the parameter space $\{X,T,P\}$ and introduce the general form of the Landau free energy $L$. With the help of the Landau free energy, we can understand the transition behavior between different states of the van der Waals system and black hole thermodynamic system in a more intuitive and natural way. Especially for the charged AdS black hole system, we analyze its Landau free energy analytically. The real state is located at the extreme point of Landau free energy. Its global minimum corresponds to the stable state of the black hole, its local minimum corresponds to the metastable state of the black hole, and its local maximum corresponds to the unstable state of the black hole. More importantly, we point out that the splitting of the global minimum of Landau free energy reflects the continuous phase transition (second-order phase transition) of a thermodynamic system, and the transformation between the global minimum of free energy reflects the first-order phase transition.

Looking at the analysis scheme of black hole thermodynamics, there are three equivalent methods, which are graphically shown in Fig.~\ref{fig4}.
\begin{itemize}
  \item In the $(T, S)$ plane (or $(P, V)$ plane), the oscillation behavior in the equation of state of the black hole is very similar to that of the van der Waals fluid. We use Maxwell equal area law to eliminate this unstable oscillation, and make the black hole from one state to another through the isobaric process (or isothermal process), so as to analyze the phase transition behavior of the black hole.
  \item The typical swallow tail structure of Gibbs free energy just corresponds to the oscillation behavior in the equation of state. The thermodynamic system always tends to the state of lower Gibbs free energy. The intersection point of swallow tail is exactly the isobaric (or isothermal) process in Maxwell equal area law, while the tip points of swallow tail structure correspond to two local extreme points in the equation of state.
  \item Landau free energy reflects all the states of the simple thermodynamic system. The extreme point of Landau free energy represents the real state of the black hole we studied. The splitting of the global minimum reflects the second-order phase transition of the black hole, while the transformation of the global minimum reflects the first-order phase transition of the black hole.
\end{itemize}

In a word, the view of Landau free energy can reflect the phase transition process of black hole more comprehensively and intuitively. This is similar to the treatment of free energy landscape, in which the temperature is a free variable, or the temperature of the ensemble. In fact, both the equation of state and the Gibbs free energy are equivalent in describing the phase transition of a thermodynamic system. When we relax a certain parameter of these two quantities, just like replacing the volume in the equation of state with the variable $X$ in this paper, or replacing the temperature of thermodynamic system in the Gibbs free energy with temperature of the ensemble, we can realize the re-understanding of the phase transition by some operation, such as the integral treatment~(\ref{landau}) we use, or the free energy landscape in~\cite{Li2020a,Li2020b,Wei2020b,Li2020c}.

In~\cite{Bhattacharya2017}, based on Landau theory of continuous phase transition (or the usual second-order phase transition) and a standard construction scheme for the Landau function in general statistical mechanics, authors give a general expression for the Landau function near the critical point of the second-order phase transition (van der Waals-type phase transition) for AdS black holes, which can correctly reproduce the known critical exponents. Then, the scheme, which is in a metric-independent way, is embodied in the charged AdS black hole in~\cite{Guo2019}, and the Landau function of the charged AdS black hole near the critical point of the van der Waals-type phase transition is presented in detail.

In our present work, we do not construct the Landau function from the critical point of the second-order phase transition, but construct the so-called Landau functional by using the idea of the variational principle. We consider a simple thermodynamic system. When it is in equilibrium, we can get its equation of state, which is an equation satisfied by three physical quantities: pressure $P$, volume $V$ and temperature $T$. Under isothermal condition, there will be transition between different phases with the change of pressure, or in the isobaric case, with the change of temperature, the transition will occur between different phases. It can be seen that the volume is the best response parameter for labeling different phases. The process of a system from an unknown state (or non-equilibrium state) to an equilibrium state can be understood mathematically as: selecting a truly stable relation (corresponding to the equation of state in equilibrium) from all possible functional relations satisfied by three physical quantities ($P$, $V$ and $T$), that is, summing (integrating) all possible functional relations to get a functional, then the truly stable relation is to make the functional take a minimum. The functional constructed in this way can directly reflect various phase transitions of the system, such as the second-order phase transition and first-order phase transition shown in Figs.~\ref{fig2} and~\ref{fig3}.

The present scheme is not only suitable for describing van der Waals-type phase transition, but also we think that this approach can be applied to arbitrary black hole thermodynamic system with the AdS background, like Hawking-Page phase transition~\cite{Hawking1983} and even reentrant phase transitions~\cite{Zangeneh2018}, from which we can read more interesting information contained in black holes. For general statistical physics system, variables $P$, $T$, $S$ and $V$ are independent of each other, and the relationship between them constitutes the equation of state and other response functions. Therefore, the present method can also be extended to general statistical physics, but the key is how to get the equation of state of the system.

In the following analysis, we plan to discuss these aspects.
     \begin{itemize}
       \item From some basic principles, we want to understand the effect of boundary terms and boundary conditions in the current Landau functional.
       \item We need to analyze the Landau functional of different black hole systems, so as to give some intuitive understanding of more complex phase transitions.
       \item It is more important to analyze the symmetry of the Landau functional and connect it with some thermodynamic properties of the system. In addition, the critical exponent of phase transition can be analyzed by means of Landau functional analysis. Meanwhile we also believe that the Landau functional obtained in this way can also be used to correctly reproduce the known critical exponents.
       \item We can try to use the stochastic process (like the Fokker-Planck equation in non-equilibrium statistical physics) to analyze the dynamics of phase transition, in order to obtain more details of the phase transition process from the perspective of kinetics.
     \end{itemize}
In the process of follow-up research, some problems may be encountered here are as follows:
  \begin{itemize}
    \item How to obtain the equation of state of black hole system, especially for the black hole without AdS background. How to define its thermodynamic volume and the corresponding equation of state, is still a problem in the research of black hole thermodynamics.
    \item For the case where the expression of thermodynamic volume is more complex, it is a technical problem that how to obtain the Landau functional accurately.
  \end{itemize}
All of these are very meaningful topic, and we will discuss them in detail in the future.

\section*{Acknowledgments}
Project funded by China Postdoctoral Science Foundation (Grant No. 2020M673460), National Natural Science Foundation of China (Grant Nos. 11947208 and 12047502), Scientific Research Program Funded by Shaanxi Provincial Science and Technology Department (2019JQ-081). This research is supported by The Double First-class University Construction Project of Northwest University. The authors would like to thank the anonymous referees for the helpful comments that improve this work greatly.

\end{document}